\documentclass[twocolumn,prl,aps,epsfig]{revtex4}
\usepackage{psfig}
\def\beq{\begin{equation}}
\def\eeq{\end{equation}}
\def\beqa{\begin{eqnarray}}
\def\eeqa{\end{eqnarray}}
\def\MeV{\nobreak\,\mbox{MeV}}
\def\GeV{\nobreak\,\mbox{GeV}}

\def\pli{p^\prime}
\def\mli{{M^\prime}^2}

\begin{document}

\title{The $J/\psi DD^*$ vertex}

\author{R. Rodrigues da Silva, R.D. Matheus, F.S. Navarra and M. Nielsen}
\affiliation{Instituto de F\'{\i}sica, Universidade de S\~{a}o Paulo\\
 C.P. 66318,  05315-970 S\~{a}o Paulo, SP, Brazil}

\begin{abstract}
We employ  QCD sum rules to calculate the $J/\psi DD^*$ form factors and 
coupling constant by studying the three-point $J/\psi D^*D$ correlation
function. We find that the momentum dependence of the form factor depends 
on the off-shell meson. We get a value for the coupling which is in agreement
with estimates based on constituent quark model.

\end{abstract}


\maketitle


Hadrons are composites of the underlying quarks whose effective fields
describe point-like physics only when all the interacting particles are
on mass-shell. When at least one of the particles in a vertex is off-shell,
the finite size effects of the hadrons become important. Therefore, the
knowledge of the form factors in hadronic vertices is of crucial importance
to estimate any hadronic amplitude using hadronic degrees of freedom. 
This work is devoted to the study of the $J/\psi D^*D$ form factor, which
is important, for instance, in the evaluation of the  dissociation cross 
section of $J/\psi$ by pions and $\rho$ mesons using effective 
Lagrangians \cite{osl,nnr,haga2}.
Since a decrease of $J/\psi$ production in heavy ions collisions
might signal the formation of a quark-gluon plasma
(QGP) \cite{ma86}, a precise evaluation of the background, i.e.,  conventional 
$J/\psi$ absorption by co-moving  pions and $\rho$ mesons, is of fundamental 
importance.

The $J/\psi D^*D$  coupling has been studied by some authors 
using different approaches: vector meson dominance model plus relativistic
potential model \cite{osl} and constituent  quark meson model \cite{dea}.
Unfortunately, the numerical results from these 
calculations may differ by almost a factor two. The relevance of this 
difference can not be underestimated since the cross section is proportional
to the square of the coupling constants. In ref.~\cite{haga2} it was shown
that the use of different coupling constants and form factors can lead to cross
sections that differ by more than one order of magnitude, and that can even 
have a different behavior as a function of $\sqrt{s}$. 

In previous works we have used the QCD sum rules (QCDSR) to
study the $D^*D\pi$ \cite{nos1, nos2}, $DD\rho$ 
\cite{nos3} and $J/\psi DD$ \cite{nos4} form factors, considering two
different mesons off mass-shell. In these works the QCDSR results for 
the form factors were parametrized by analytical forms
such that the respective extrapolations at the off-shell meson poles
provided consistent values for the corresponding coupling constant.
In this work we use the QCDSR approach to evaluate the 
$J/\psi D^*D$  form factors and use the same procedure described above to
estimate the $J/\psi D^*D$ coupling constant. 

\begin{figure}
\centerline{\psfig{file=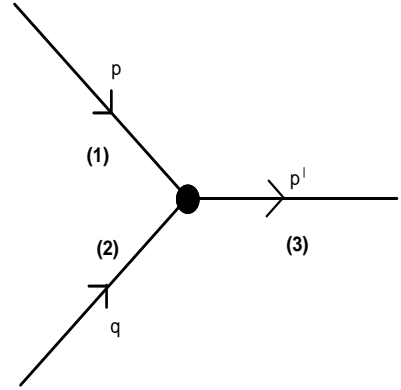,height=2in,width=2in,angle=0}}
\caption{Diagram representing the $H_1(p)H_2(q)H_3(\pli)$ vertex.}
\label{diagrams}
\end{figure}

The three-point function associated with a $H_1H_2H_3$ vertex (see fig.~1), 
where $H_1$ and
$H_3$ are the incoming and outgoing external mesons respectively and $H_2$
is the off-shell meson, is given by
\beqa
\Gamma_{\mu\nu}(p,\pli)&=&\int d^4x \, d^4y \,   
\, e^{ip^\prime.x} \, e^{-i(\pli-p).y}\nonumber\\ 
&\times&\langle 0|T\{j_3(x)
j^\dagger_2(y)j^\dagger_1(0)\}|0\rangle\;,
\label{cor}
\eeqa
where $j_i$ is the interpolating field for $H_i$. For $J/\psi$, $D^*$
and $D$ mesons the interpolating fields are respectively $j_\mu^{(\psi)}=
\bar{c}\gamma_\mu c$, $j_\nu^{(D^*)}=\bar{q}\gamma_\nu c$ and
$j^{(D)}=i\bar{q}\gamma_5 c$  with $q$ and $c$ being a light quark and 
the charm quark fields.

The phenomenological side of the vertex function, $\Gamma(p,p^\prime)$,
is obtained by the consideration of $H_1$ and $H_3$ state contribution to
the matrix element in Eq.~(\ref{cor}):
\beqa
\Gamma_{\mu\nu}^{(phen)}(p,\pli)={1\over p^2-m_{1}^2}{1\over{\pli}^2-m_3^2}
\langle 0|j_3|H_3(\pli)\rangle
\nonumber \\*[7.2pt]
\times \langle H_3(\pli)|j^\dagger_2|H_1(p)\rangle\langle H_1(p)|
j^\dagger_1|0\rangle +\mbox{h.~r.}\; ,
\label{phe1}
\eeqa
where h.~r. means higher resonances.

The matrix element of the current $j_2$ defines the
vertex function $V_{\lambda\lambda'}(p,\pli)$:
\beq
\langle H_3(\pli)|j^\dagger_2|H_1(p)\rangle=\langle H_2(q)|j^\dagger_2|0
\rangle{V_{\lambda\lambda'}(p,\pli)\over q^2-m_2^2}\; ,
\label{for}
\eeq
where $q=\pli-p$. Calling $p_1,~p_2$ and $p_3$ the four momentum of
$J/\psi,~D^*$ and $D$ respectively one has
\beq
V^{\lambda\lambda'}(p_1,p_2,p_3)=g_{\psi DD*}(q^2)
\epsilon^{\alpha\beta\gamma\delta}\epsilon_\alpha^\lambda(p_1)
\epsilon_\gamma^{\lambda'}(p_2)p_{3\beta}p_{2\delta}\;.
\label{ver}
\eeq
 The vacuum to meson 
transition amplitudes appearing in Eqs.~(\ref{phe1}) and (\ref{for})
are given in terms of
the corresponding meson decay constants $f_{H_i}$ by
\beq
\langle 0|j^{(D)}|D\rangle={m_D^2f_{D}\over m_c}\;,
\label{fd}
\eeq
and
\beq
\langle V(p,\epsilon)|j^\dagger_\alpha|0\rangle=m_{V}f_{V}\epsilon^*_\alpha
\; ,
\label{fv}
\eeq
for the vector meson $V=J/\psi$ or $V=D^*$.
Therefore, using Eqs.~(\ref{for}), (\ref{ver}),  (\ref{fd}) and (\ref{fv}) in
Eq.~(\ref{phe1}) we get
\beqa
\Gamma_{\mu\nu}^{(phen)}(p,\pli)&=&C{g_{\psi DD*}(q^2)
\epsilon_{\alpha\beta
\mu\nu}p^\alpha{\pli}^\beta\over (q^2-m_2^2)(p^2-m_1^2)({\pli}^2-m_3^2)}
\nonumber\\
&+& \mbox{h.~r.}\; ,
\label{phen}
\eeqa
where
\beq
C={m_D^2m_{D^*}m_{\psi}f_Df_{D^*}f_{\psi}\over m_c}\; .
\label{chh}
\eeq
The contribution of higher resonances and continuum in Eq.~(\ref{phen})
will be taken into account as usual in the standard form of 
ref.~\cite{io2}.

The QCD side, or theoretical side, of the vertex function is evaluated by
performing Wilson's operator product expansion (OPE) of the operator
in Eq.~(\ref{cor}). Writing $\Gamma_{\mu\nu}$ in terms of the invariant
amplitude:
\beq
\Gamma_{\mu\nu}(p,\pli)=\Lambda(p^2,{\pli}^2,q^2)\epsilon_{\alpha\beta
\mu\nu}p^\alpha{\pli}^\beta\;,
\eeq
we can write a double dispersion relation for $\Lambda$,
over the virtualities $p^2$ and ${\pli}^2$
holding $Q^2=-q^2$ fixed:
\beq
\Lambda(p^2,{\pli}^2,Q^2)=-{1\over4\pi^2}\int ds
du~ {\rho(s,u,Q^2)\over(s-p^2)(u-{\pli}^2)}\;,
\label{dis}
\eeq
where $\rho(s,u,Q^2)$ equals the double discontinuity of the amplitude
$\Lambda(p^2,{\pli}^2,Q^2)$ on the cuts $s_{min}\leq s\leq\infty$,
$m_c^2\leq u\leq\infty$, with $s_{min}=4m_c^2$ in the case of off-shell 
$D^*$ or $D$ and $s_{min}=m_c^2$ in the case of off-shell $J/\psi$.
We consider diagrams up to dimension three which include the perturbative 
diagram and the quark condensate. 
To improve the matching between the two sides 
of the sum rules, we perform a double Borel transformation  in both variables
$P^2=-p^2\rightarrow M^2$ and ${P^\prime}^2=-{\pli}^2\rightarrow\mli$ 
We get one sum rule for each meson considered off-shell. Calling 
$g_{\psi DD^*}^M(q^2)$ the $\psi DD^*$ form factor for the off-shell meson $M$,
we get the following sum rules:
\begin{widetext}
\beq
C~\frac{g_{\psi DD^{*}}^{(D)}(t)}{(t- m_{D}^2)}
e^{-\frac{m_{D^{*}}^2}{\mli}}
e^{-\frac{m_{\psi}^2}{M^2}}
=\frac{1}{4\pi^2}
\int_{4m^2}^{s_0}
\int_{u_{min}}^{u_0}
dsdu
\rho^{(D)}(u,s,t)e^{-\frac{s}{M^{2}}}e^{-\frac{u}{\mli}}\Theta(u_{max}-u),
\label{ffd}
\eeq
\beq
C~\frac{g_{\psi DD^{*}}^{(D^*)}(t)}{(t- m_{D^{*}}^2)}
e^{-\frac{m_{D}^2}{\mli}}
e^{-\frac{m_{\psi}^2}{M^2}}
=\frac{1}{4\pi^2}
\int_{4m^2}^{s_0}
\int_{u_{min}}^{u_0}
dsdu
\rho^{(D^*)}(u,s,t)e^{-\frac{s}{M^{2}}}e^{-\frac{u}{\mli}}\Theta(u_{max}-u),
\label{ffds}
\eeq
and
\beq
C~\frac{g_{\psi DD^{*}}^{(J/\psi)}(t)}{(t- m_{\psi}^2)}
e^{-\frac{m_{D}^2}\mli}
e^{-\frac{m_{D^{*}}^2}{M^2}}
=\frac{1}{4\pi^2}
\int_{m^2}^{s_0}
\int_{u_{min}}^{u_0}
dsdu
\rho^{(J/\psi)}(u,s,t)e^{-\frac{s}{M^{2}}}e^{-\frac{u}{\mli}}\Theta(u_{max}-u),
\label{ffpsi}
\eeq
\end{widetext}
with $t=q^2$,
\beq
\rho^{(D)}(u,s,t)=\rho^{(D^*)}(u,s,t)=
\frac{3m_c}{\sqrt{\lambda}}
\left(
1+\frac{s\lambda_2}{\lambda}
\right),
\eeq
$\lambda =(u+s-t)^2 -4us$, $\lambda_2 =u+t-s +2m_c^2$ and
\begin{widetext}
\beq
u^{max}_{min}=\frac{1}{2m_c^2}
\left[
-st+ m_c^2(s+ 2t) 
\pm 
\sqrt{s(s-4m_c^2)(t-m_c^2)^2}
\right],
\eeq
in the case of off-shell $D$ or $D^*$. In the case of an off-shell $J/\psi$
we get:
\beqa
\rho^{(J/\psi)}(u,s,t)=
\frac{3m_c}{\lambda^{3/2}}
\left[
(u-s)^2 -t(u+ s- 2m_c^2)
\right]-4\pi^2<\bar{q}q>\delta(s-m_c^2)\delta(u-m_c^2),
\eeqa
and
\beq
u^{max}_{min}=\frac{1}{2m_c^2}
\left[
-st+ m_c^2(2s+ t) 
\pm 
\sqrt{t(t-4m_c^2)(s-m_c^2)^2}
\right].
\eeq
\end{widetext}

In the Eqs.~(\ref{ffd}), (\ref{ffds}) and (\ref{ffpsi}) we have 
transferred to the QCD side the higher resonances contributions 
through the introduction of
the continuum thresholds $s_0$ and $u_0$.

\begin{figure}[htb]
\centerline{\psfig{figure=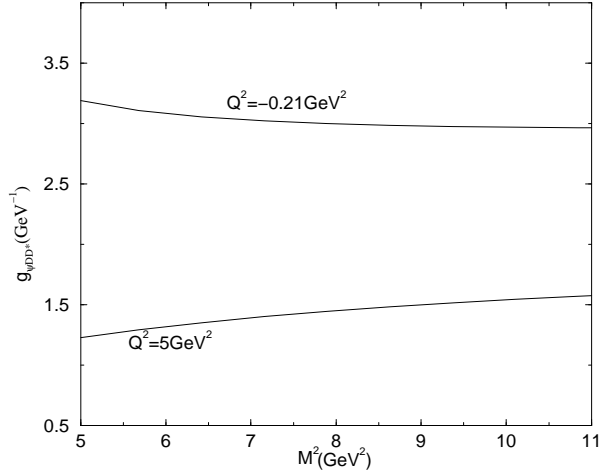,width=8cm,angle=0}}
\protect\caption{$M^2$ dependence of $g_{J/\psi DD^*}^{(D)}(Q^2)$ for
$Q^2=-0.21\,\GeV^2$ and $Q^2=5.0\,\GeV^2$.}
\label{fig2}
\end{figure}

The parameter values used in all calculations are $m_c=1.3\,
\GeV$, $m_D=1.87\,\GeV$, $m_{D^*}=2.01\,
\GeV$, $m_\psi=3.1\,\GeV$, $f_{D}=(170 \pm 10)~\MeV $,
$f_{D^{*}}=(240 \pm 20)~\MeV $, $f_{J/\psi}=(405 \pm 15)\MeV $,
$\langle\overline{q}q\rangle\,=\,-(0.23)^3\,\GeV^3$. 
The continuum thresholds for the sum rules are $s_0=(m_{1}+\Delta_s)^2$ 
and $u_0=(m_{3}+\Delta_u)^2$
with $\Delta_s=\Delta_u=0.5\GeV$.

We first discuss the $J/\psi DD^*$ form factor with an off-shell $D$ meson. 
In Fig.~2 we
show the behavior of the form factor $g_{J/\psi DD^*}^{(D)}(Q^2)$ 
at $Q^2=5.0\,\GeV^2$ and $Q^2=-0.21~\GeV^2$, as a function
of the Borel mass $M^2$ using $\mli=M^2{m_{D^*}^2\over m_\psi^2}$.
We can see that the QCDSR results are rather stable
in the interval $7\leq M^2\leq11\,\GeV^2$. In Fig.~3 we show 
$g_{J/\psi DD^*}^{(D)}(Q^2=-0.21\GeV^2)$ as a function of $M^2$ and $\mli$.

\begin{figure}[htb]
\centerline{\psfig{figure=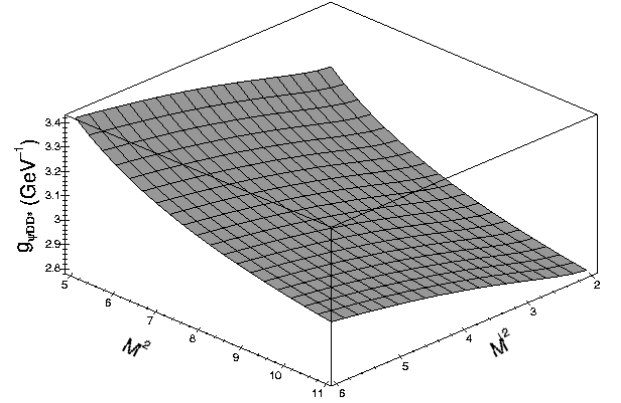,width=8cm,angle=0}}
\protect\caption{$M^2$ and $\mli$ dependence of 
$g_{\psi DD^*}^{(D)}(Q^2=-0.21\GeV^2)$ .}
\label{fig3}
\end{figure}
From Fig.~3 we see that the stability is still good even considering the two
independent Borel parameters. The same kind of stability is obtained for 
other values of $Q^2$ and for the other two form factors.

Fixing $M^2=m_1^2$ and $\mli=m_3^2$ we  show, 
in Fig.~4, the momentum dependence of the QCDSR results for the form factors 
$g_{\psi DD^*}^{(D)}$, $g_{\psi DD^*}^{(D^*)}$ and $g_{\psi DD^*}^{
(J/\psi)}$ through the circles, squares and triangles respectively.
Since the present approach 
cannot be used at $Q^2<<0$, to extract the $g_{\psi DD^*}$
coupling from the form factors we need to extrapolate the curve to 
$Q^2=-m_2^2$: the mass of the off-shell meson.

\begin{figure}[htb]
\centerline{\psfig{figure=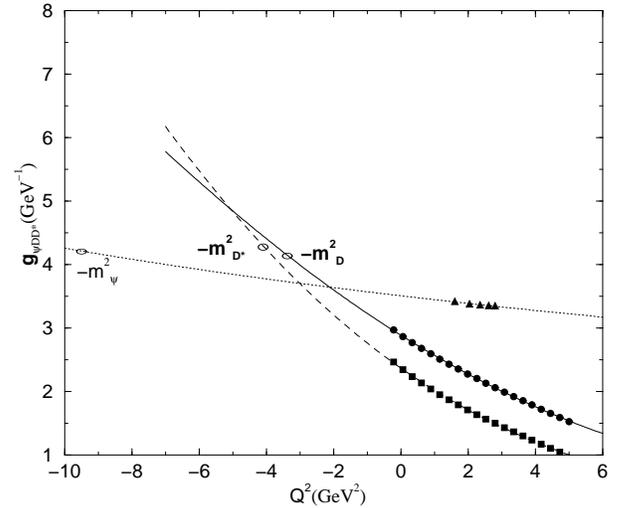,width=8cm,angle=0}}
\protect\caption{Momentum dependence of the $J/\psi DD^*$ form factors. The 
dotted, dashed and solid lines give the 
parameterization of the QCDSR results (triangles, squares and circles) through 
Eqs.~(\protect\ref{papsi}),
(\protect\ref{pads}) and (\protect\ref{pad}) respectively.}
\label{fig4}
\end{figure}
In order to do this extrapolation we fit 
the QCDSR results  with an analytical expression. We tried to fit
our results with a mono-pole form, since this is very often used 
for form factors, but the fit was only good for $g_{\psi DD^*}^{(J/\psi)}$. 
For $g_{\psi DD^*}^{(D)}$ and $g_{\psi DD^*}^{(D^*)}$
we obtained  good fits using  a Gaussian form. We get:
\beq
g^{(J/\psi)}_{\psi DD^{*}}(Q^2)=
\frac{199.2}{Q^{2}+ 56.8},
\label{papsi}
\eeq
\beq
g^{(D^{*})}_{\psi DD^{*}}(Q^2)=
19.9
\mbox{exp}\left[
-\frac{(Q^2+27)^{2}}{345}
\right],
\label{pads}
\eeq
\beq
g^{(D)}_{\psi DD^{*}}(Q^2)=
12.7
\mbox{exp}\left[
-\frac{(Q^2+25.8)^{2}}{450}
\right].
\label{pad}
\eeq
These fits are also shown in Fig.~4 through the dotted, dashed and solid 
lines respectively. From Fig.~4 we see that all three form factors lead
to compatible values for the coupling constant when the form factors are 
extrapolated to the off-shell meson mass (shown as open circles in Fig.~4).
 Considering the uncertainties in  
the continuum threshold, and the difference 
in the values of the coupling constants extracted when the $D$, $D^*$ or 
$J/\psi$ mesons
are off-shell, our result for the $J/\psi DD^*$ coupling constant is:

\beq
g_{\psi DD^{*}}=(3.48 \pm 0.76)\mbox{GeV}^{-1}.
\label{fi}
\eeq

In Table I we show the results obtained for the same coupling constant 
using different approaches.in 
refs.~\cite{osl} and \cite{dea}.

\begin{center}
\begin{tabular}{|c|c|c|}
\hline
&&\\
this work & ref.~\cite{osl} & ref.~\cite{dea}\\
\hline\hline
 3.48 $\pm$ 0.76 & 8.02 $\pm$ 0.62& 4.05 $\pm$ 0.25\\
\hline 
\end{tabular}
\end{center}
\begin{center}
{\small{\bf TABLE I:}  Values of the coupling constant $g_{\psi DD^{*}}$
in $\GeV^{-1}$ evaluated using different approaches.}
\end{center}

While our result is compatible with the coupling obtained using
constituent  quark meson model \cite{dea}, it is half of the value
obtained with the vector meson dominance model plus relativistic
potential model \cite{osl}.

To summarize: we have used the method of QCD sum rules  to compute form factors
and coupling constant in the $J/\psi D D^*$ vertex. Our results for the 
coupling show once more 
that this method is robust, yielding numbers which are approximately the 
same regardless of which particle we choose to be off-shell and depending 
weakly on the choice of the continuum threshold. As for the form factors,
we obtain a harder (softer) form factor when the off-shell particle is 
heavier (lighter).

\section*{Acknowledgments}

This work was supported by CNPq and FAPESP.

\end{document}